\documentclass[10pt]{iopart}

\usepackage[utf8]{inputenc}

\usepackage{graphicx}
\usepackage{tikz}
\usepackage{amssymb}
\usepackage{braket}
\usepackage[hidelinks]{hyperref}
\usepackage{printlen}
\usepackage{epstopdf}

\usepackage{hhline}

\usepackage{cite}

\begin{document}
\title[\small Interplay between van der Waals and dipole--dipole interactions among Rydberg atoms]{Interplay between van der Waals and dipole--dipole interactions among Rydberg atoms}

\author{Julius de Hond$^{1,2}$,
Nataly Cisternas$^1$\footnote{Present address: Laboratorio de Innovaci\'on e Investigaci\'on Educativa, IDEClab, Universidad de Concepci\'on, Concepci\'on, Chile},
R.\ J.\ C.\ Spreeuw$^1$, H.\ B.\ van Linden van den Heuvell$^1$, N.\ J.\ van Druten$^1$\footnote{Corresponding author: n.j.vandruten@uva.nl}}

\address{$^1$ Van der Waals--Zeeman Institute, Institute of Physics, University of Amsterdam, Science Park 904, 1098 XH Amsterdam, The Netherlands}
\address{$^2$ Research Laboratory of Electronics, MIT--Harvard Center for Ultracold Atoms, and Department of Physics, Massachusetts Institute of Technology, Cambridge, MA 02139, USA}

\begin{abstract}
Coherently manipulating Rydberg atoms in mesoscopic systems has proven challenging due to the unwanted population of nearby Rydberg levels by black-body radiation. Recently, there have been some efforts towards understanding these effects using states with a low principal quantum number that only have resonant dipole--dipole interactions. We perform experiments that exhibit black-body-induced dipole--dipole interactions for a state that also has a significant van der Waals interaction. Using an enhanced rate-equation model that captures some of the long-range properties of the dipolar interaction, we show that the initial degree of Rydberg excitation is dominated by the van der Waals interaction, while the observed linewidth at later times is dominated by the dipole--dipole interaction. We also point out some prospects for quantum simulation.
\end{abstract}	

\submitto{\jpb}
\maketitle

Rydberg atoms have been celebrated for their strong, tunable interactions that make them interesting for applications in quantum information and simulation. While there has been significant progress with a few Rydberg atoms in arrays of optical tweezers \cite{Barredo14,Ravets14,Bernien17} and in small ensembles \cite{Ebert15}, the progress obtained with larger numbers of atoms has been more modest. Rydberg dressing of Bose--Einstein condensates, for instance, has proven difficult \cite{Balewski14-2,Plodzien16}, and typically there are a handful of mechanisms at work limiting the coherence.

One endemic problem is the undesired decay to nearby Rydberg states under the influence of black-body radiation (BBR). These states have a large, resonant dipole--dipole interaction with the state being excited primarily, leading to rapid dephasing of the coherent excitation \cite{Goldschmidt16,Boulier17,Young17}. While this effect has been known for some time \cite{Gallagher79}, the implications have only recently been recognized in a quantum information context. A solution is to operate experiments under cryogenic conditions \cite{Hermann14,Boulier17}, and to completely squeeze out any remaining BBR by operating in the narrow gap between two conducting parallel plates \cite{Nguyen18}. While these are interesting approaches, they require larger expenses and increase the complexity of setups. Most experiments will have to cope with these undesired pollutant atoms, and the dephasing they impose.

\begin{figure}[t]
	\centering
	\includegraphics{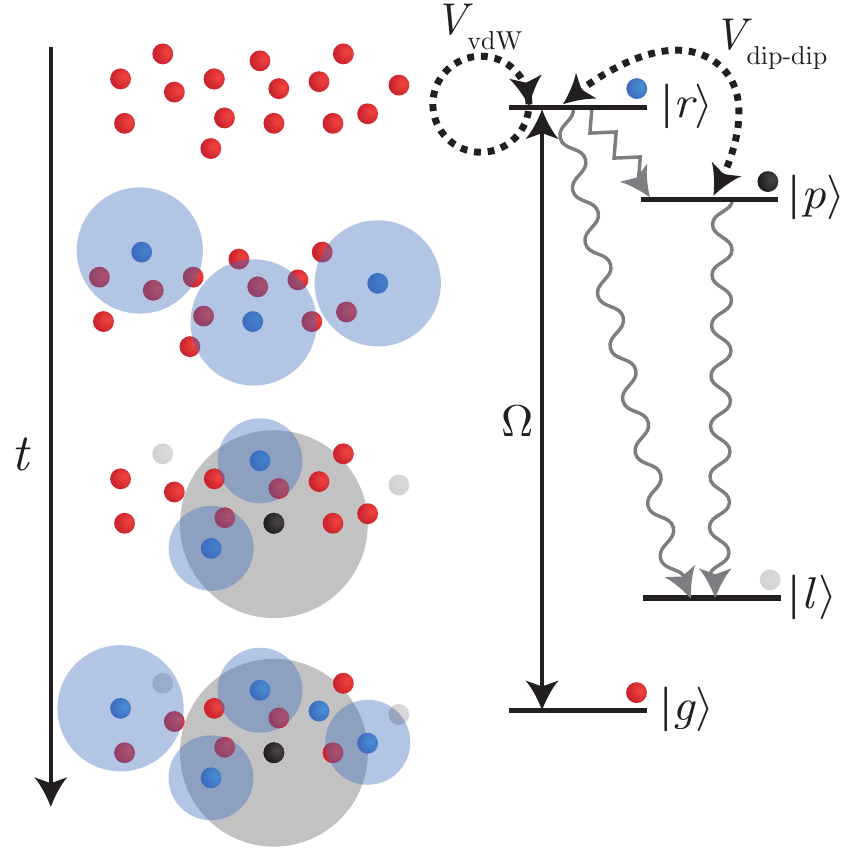}
	\caption{Left: illustration of the dynamics in our experiment. Ground state atoms (red) get excited to Rydberg states (blue), separated by the range of the van der Waals blockade (blue disks). Rydberg states either decay out of the system (gray) or, due to a black-body photon, to a nearby Rydberg state (black). This pollutant state strongly limits the coherence in its vicinity (gray disks), thereby reducing the blockade radii of the Rydberg atoms surrounding it.
	 Right: single-atom level diagram of our experiment. The straight arrow denotes the coherent driving by the laser light, the wavy and jagged arrows denote spontaneous and BBR-stimulated decay, respectively; the dashed arrows indicate both the van der Waals and the dipolar exchange interaction.}
	\label{fig:level-diagram}
\end{figure}

The essence of the problem lies in that the pollutant states that are populated via BBR have lifetimes that are similar to or longer than that of the original state, which means atoms can get shelved there. 
By the very nature of their creation mechanism, these shelved atoms have a strong, resonant dipole--dipole exchange interaction with atoms in the initial Rydberg state. This interaction is usually much stronger than all other energy and frequency scales involved, and leads to dephasing, causing significant broadening of the Rydberg linewidth.

Here, we present experiments in an ultracold gas where the BBR-induced dipole--dipole interaction is in strong competition with the van der Waals interaction among excited Rydberg atoms. Under conditions where the latter would limit the degree of excitation (`blockade'), we show that such van-der-Waals-interacting systems are extremely sensitive to (BBR-induced) impurity Rydberg states, down to the level of only a few impurities within the cloud (figure~\ref{fig:level-diagram}). This combination has not been studied experimentally before, and leads to different behavior compared to previous experiments. We find that the energy scale of the dipolar exchange is comparable to that of the van der Waals interaction under our experimental conditions.

We describe the interplay of these two types of interaction among $N\approx 10^4$ atoms with a relatively simple model based on rate equations. The model, with fully \emph{a priori} determined atomic parameters, describes our data reasonably well. Furthermore, we point out that accounting for BBR-induced impurities naturally leads to interesting extensions of existing Rydberg-based proposals; for instance models for epidemic dynamics \cite{Perez-Espigares17,Grassberger83} would now allow for including a generalized `incubation period' \cite{Li99}. This highlights the importance of BBR for Rydberg-based interactions, and shows that BBR should not only be regarded as a limiting factor, but may actually be an important contributor to realizing interesting (quantum) many-body systems with Rydberg atoms.

We start by describing the experimental results. The setup is very similar to that of our recent work on the coherent collective onset and Rydberg blockade \cite{deHond18} on an atom chip \cite{vanEs10}. Here we focus on longer timescales where BBR-induced Rydberg--Rydberg transitions also become relevant. In short, the starting point is  a cloud of approximately $2\times 10^4$ $^{87}$Rb atoms in the $\ket{F=2, m_F=2}$ Zeeman sublevel of the electronic ground state, at a temperature of $2~\mathrm{\mu K}$, magnetically trapped in vacuum about $100~\mathrm{\mu m}$ below the gold surface of the chip. The trap is cigar-shaped and has axial and radial trapping frequencies of $\omega_\parallel/2\pi = 46~\mathrm{Hz}$ and $\omega_\perp/2\pi = 860~\mathrm{Hz}$, respectively. Under these conditions the cloud has a peak density $n_0\approx 4\,\mu$m$^{-3}$.
\begin{figure}
	\centering
	\includegraphics{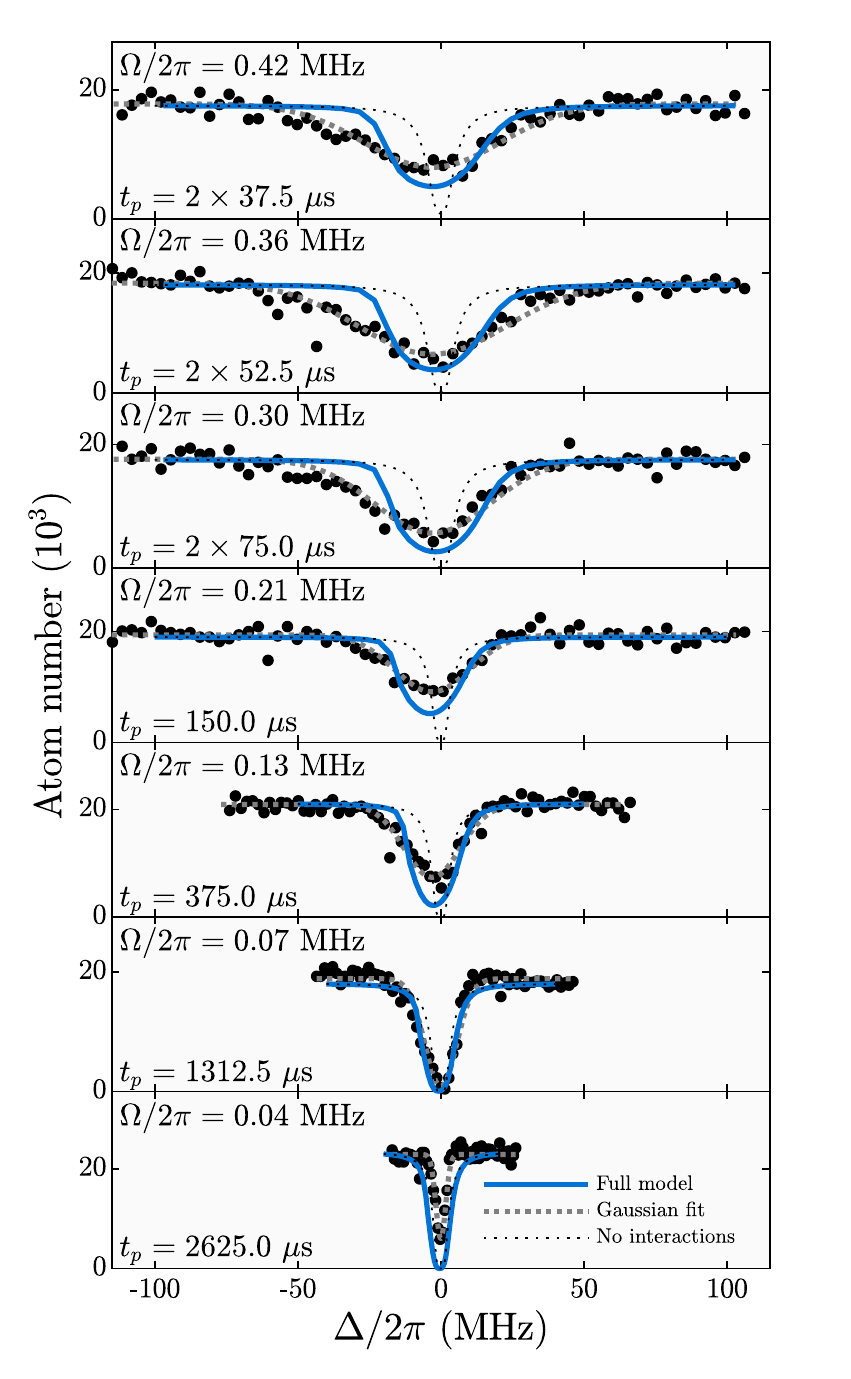}
	\caption{Measured $\ket{28D_{5/2},m_J=5/2}$ spectra for varying Rabi frequency $\Omega$ (points), and Gaussian functions fitted to the data (dashed, gray lines). The solutions to the cellular model with added pollutant states are shown by the blue, solid lines, these show a clear deviation away from the non-interacting case (dotted, black lines). In the experiments we varied the number of pulses and the pulse time $t_p$ to obtain a strong enough loss signal, both these parameters and the Rabi frequency are included in each plot.}
	\label{fig:cascade-plot}
\end{figure}

For  Rydberg excitation we use a two-photon excitation scheme with counter-propagating lasers with 480 and $780~\mathrm{nm}$ wavelengths which are stabilized to a reference cavity \cite{deHond17}. The lasers are tuned near the  $\ket{28D_{5/2},m_J=5/2}$ Rydberg level, with a  van der Waals interaction coefficient $C_6/2\pi = 21.5~\mathrm{MHz~\mu m^6}$ (root-mean-square angle-averaged; with positive $C_6$ corresponding to attractive interactions: $V_\mathrm{vdW}=-C_6/r^6$) and lifetime $\tau_r = 15.2~\mathrm{\mu s}$ \cite{Sibalic17} (at a BBR temperature of $300~\mathrm{K}$). The excitation scheme uses the intermediate $5P_{3/2}$ state, with a detuning of $100~\mathrm{MHz}$ to reduce scattering off of it. 
The remaining atom number after the Rydberg excitation is detected using absorption imaging of the cloud after time of flight.

By scanning the frequency of the 480-nm laser we obtain Rydberg loss spectra, which are shown in figure~\ref{fig:cascade-plot}. For each scan we set the two-photon Rabi frequency $\Omega$ and pulse duration $t_p$ by adjusting the power and pulse time of the 780-nm laser, such that we have an observable loss signal. For short pulse times ($t_p<100~\mu$s) we measured the cumulative losses of two pulses separated by much more than the Rydberg lifetime, in order to improve the signal-to-noise ratio.
Note that the widths of these spectra can be several tens of MHz for pulse times of a few Rydberg lifetimes (upper curves), two orders of magnitude larger than what one would expect based on the two-photon Rabi frequency alone.
The duration of our pulses ranges from $37.5$ to $2625~\mathrm{\mu s}$ for the highest and lowest Rabi frequencies, respectively. These pulse times are longer than the lifetime of the $28D_{5/2}$ state, so there is enough time for pollutant states to become populated.

\begin{figure}
	\centering
	\includegraphics{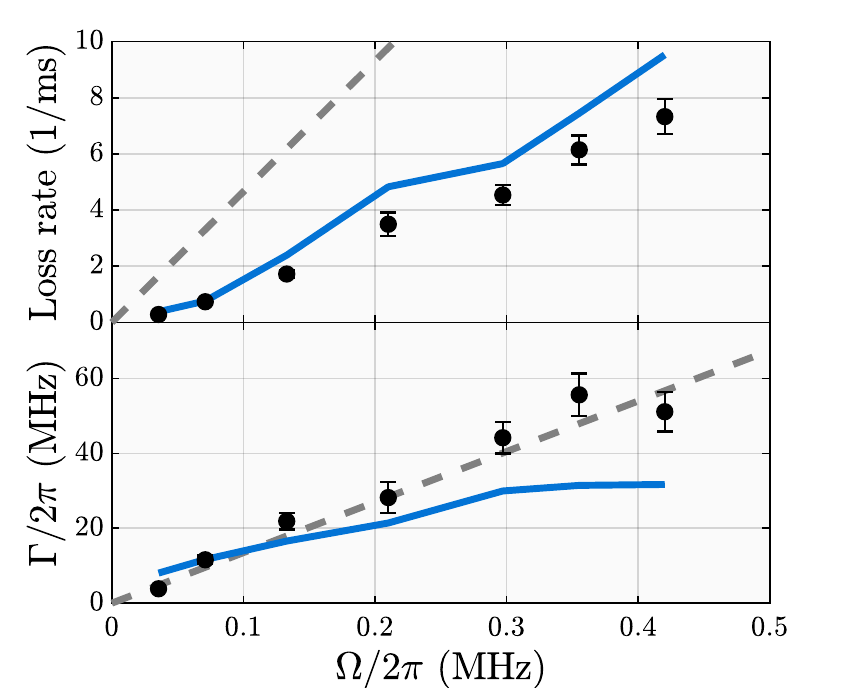}
	\caption{Linewidth $\Gamma$ and on-resonance loss rate (points) extracted from the experimental data in figure~\ref{fig:cascade-plot}, together with the steady-state scaling (i.e.\ disregarding the van der Waals interaction, dashed, gray line) and the parameters obtained from fits to spectra calculated using the cellular model (solid, blue lines). The error bars denote the 95\% confidence bound of the parameters obtained from the fits to the measured data.}
	\label{fig:linewidth-features}
\end{figure}

The observed spectra are fitted by Gaussian functions; using these fits we extract both the linewidth $\Gamma$ (full width at half maximum, FHWM) and the (average) loss rate on resonance during the excitation pulse. Both increase roughly linearly with the Rabi frequency, see figure~\ref{fig:linewidth-features}. While this is in qualitative agreement with the steady-state description of BBR-induced Rydberg populations explored previously \cite{Goldschmidt16}, a quantitative description necessitates the inclusion of van der Waals interactions, as we shall demonstrate below. Similar experiments on BBR-induced Rydberg interactions were done with the $18S_{1/2}$ state of rubidium \cite{Goldschmidt16,Boulier17,Young17}. Since the van der Waals interaction scales with the principal quantum number $n$ as $n^{11}$ \cite{Low12}, its effects were negligible in that case.

\begin{table}
	\centering
	\caption{Properties of the most important decay products $\ket{i}=\ket{nL_J}$ of the $28D_{5/2}$ state, which itself has a (BBR included) lifetime of $15.2~\mathrm{\mu s}$. All cited lifetimes $\tau_i$, branching ratios $b_i$, and transition rates include BBR at $300~\mathrm{K}$. The cited dipole--dipole interaction coefficient $C_3$ is the root-mean-square value obtained after calculating it for several angles using \cite{Sibalic17}, see the main text for more details. For comparison, the $5P_{3/2}$ state is listed as well; its dipole--dipole interaction is, of course, negligible, and it does not contribute to the dephasing.}
	\label{tab:bbr-culprits}
	\vspace{\baselineskip}
	\begin{indented}
	\item[]
	\begin{tabular}{|c|c|c|c|c|}
		\hline
		$\ket{i}$ 	& 	Transition	&	$\tau_i$ & $b_i$ & $C_3/2\pi$ \\
		$nL_{J}$	&	rate ($\mathrm{s}^{-1}$)	& ($\mu\mathrm{s}$) &	ratio (\%) & ($\mathrm{MHz\,\mu m}^3$) \\	
		\hhline{|=|=|=|=|=|}
		$26F_{7/2}$	&	4\,684	&	19.8  & 7.1 & 83.8 \\
		$27F_{7/2}$	&	3\,183	&	21.9  & 4.8  & 265.6\\
		$29P_{3/2}$	&	1\,941	&	24.2 & 3.0  & 297.3\\ 	
		$30P_{3/2}$	&	3\,052	&	26.4 & 4.6  & 92.0\\
		\hline		
		$5P_{3/2}$	&	27\,445	&	0.026& 41.7	& 0.0 \\
	\hline
	\end{tabular}
	\end{indented}
\end{table}

A description of our experiments in terms of the many-body Hamiltonian that underpins this system is not feasible in practice, due to the exorbitant size of the Hilbert space it covers.
Instead, in order to analyze the experimental data, we use a model that combines several features of models described elsewhere \cite{Boulier17,Whitlock16,deHond18,Young17}. We extend the rate-equation model presented in \cite{Boulier17} to include the van der Waals interaction, and attempt to include the long-range character of the dipolar interaction in a more realistic manner. So far, this interaction has often been described as a homogeneous broadening arising from a constant density of pollutant states. However, under that assumption the integral describing the strength of this interaction would diverge. To illustrate: take a single Rydberg atom at the origin of a homogeneous sea of pollutant atoms (with density $ n_0 \rho_p$, where $n_0$ is the overall density). The energy associated with this interaction is given by:
\begin{eqnarray}\label{eq:dipdip-interaction-integral}
	\int n_0 \rho_p \frac{C_3}{r^3} \, d\mathbf{r} \propto \int_{r_\mathrm{min}}^{r_\mathrm{max}} \frac{1}{r} \, dr,
\end{eqnarray}
which diverges logarithmically for large $r_\mathrm{max}$, highlighting the importance of the long-range character of the dipolar interaction. The lower limit, $r_\mathrm{min}$ is, in practice, set by the (van der Waals) blockade. 

Instead, our model takes the spatially-varying density into account by dividing the cloud into cells with constant density. We found this was necessary, because neither the localized rate-equation model of \cite{Boulier17}, nor the Liouville--von Neumann approach we developed earlier \cite{deHond18} described our data well. The use of rate equations instead of a description in terms of coherent population dynamics is justified by observing that the excitation times we are concerned with presently ($t_p \gtrsim 10~\mathrm{\mu s}$) are far larger than the coherence times measured in the past \cite{deHond18}.

The van der Waals interaction is included as an effective detuning in our rate equations. This is motivated by the fact that it is a diagonal term in the Hamiltonian describing the light-matter coupling. The dipolar interaction is off-diagonal, meaning it leads to increased dephasing. This will be accounted for by a decoherence term that depends on the populations throughout the cloud. Effective decay rates describe population transfer towards pollutant states and untrapped states, which register as loss.

Of course, there is more than one decay channel (see table~\ref{tab:bbr-culprits}), but under the assumption that there are no interstate interactions among the decay products (i.e.\ assuming the dominant interaction is with the initially excited Rydberg state), we can model them as a single level with an effective interaction coefficient $\widetilde{C}_3$ and branching ratio $b$.
To this end, we take the average of the $C_3$ coefficients and the decay rates, weighted by the branching ratios $b_i$ and lifetimes $\tau_i$, similar to the procedure in \cite{Goldschmidt16,Boulier17}.
That is,
\begin{eqnarray}
	\widetilde{C}_3 \equiv \frac{\sum_{i} \left| C_3^{~(i)} \right| b_i \tau_i}{ \sum_{i}{b_i\tau_i}}, \nonumber
\end{eqnarray}
where the sums run over all pollutant states.
For our parameters, we find $\widetilde{C}_3/2\pi = 165~\mathrm{MHz~\mu m^3}$ and an effective lifetime $\widetilde{\tau}_p \equiv \left[ \sum_{i} \left( b_i/\tau_i \right) / \sum_{i} b_i \right]^{-1} = 22.3~\mathrm{\mu s}$. The branching ratio $b$ into this compound level is then given by the sum of the branching ratios of the levels that constitute it, which is $\sim\! 0.20$. 
Note that BBR can, in principle, also re-excite decay products back to the initial excited state. As has been pointed out elsewhere, this can increase the effective lifetime by 10\% \cite{Galvez95, Archimi19}. Since the dynamics we observe are only weakly sensitive to the absolute lifetimes we do not expect this to make a large difference, and believe our `one-way' model to be a good approximation.

\begin{figure}
	\centering
	\includegraphics{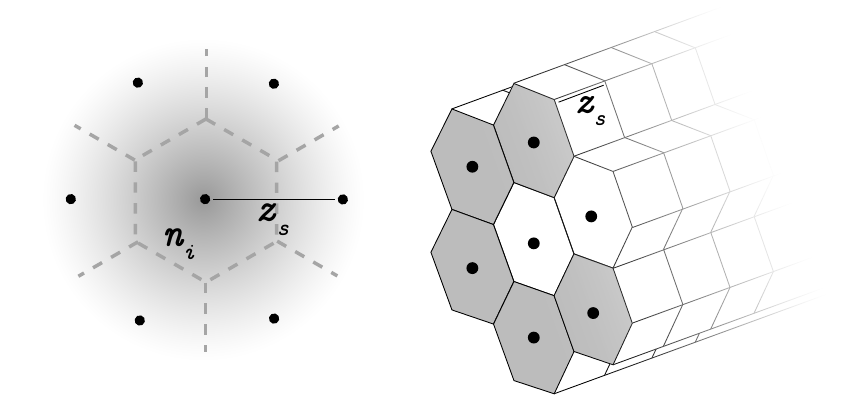}
	\caption{Illustration of the cellular method that was used to calculate the spectral properties. The dipolar interaction has a fundamentally long-range character, and thus the excitation at one point in the cloud is affected by all others. Left: transversally, we use a hexagonal grid with lattice spacing $z_s$ in which each cell is assumed to have a homogeneous density $n_i$, with associated state fractions $\rho_{g, i} \cdots \rho_{l, i}$. Right: the first two rings of the hexagonal lattice. The axial symmetry means the populations in a lot of cells will be equivalent to that in other cells (gray). We do not need to calculate these, provided we account for them in the various summations that are carried out.} 
	\label{fig:honeycomb-calculation-lattice}
\end{figure}

For a lattice consisting of cells denoted by indices $i$, these considerations result in the following rate-equation model. Here $\rho_{s,i}$ denotes the relative population in state $s \in \left\{g, r, p \right\}$ in cell $i$, $\gamma_r$ denotes the decay rate of state $\ket{r}$, and $\gamma_p = 1/\widetilde{\tau}_p$:

\begin{eqnarray}
	\dot{\rho}_{g,i} &= -\frac{\Omega^2 \gamma_i}{\gamma_i^{~2} + 4\left[\Delta + \left( 4\pi/3\right)^2 C_6 \left( n_i \rho_{r,i} \right)^2 \right]^2} \left( \rho_{g,i} - \rho_{r,i} \right) \nonumber \\
	\dot{\rho}_{r,i} &= \frac{\Omega^2 \gamma_i}{\gamma_i^{~2} + 4\left[\Delta + \left( 4\pi/3\right)^2 C_6 \left( n_i \rho_{r,i} \right)^2 \right]^2} \left( \rho_{g,i} - \rho_{r,i} \right) - \gamma_r\rho_{r,i} \nonumber \\
	\dot{\rho}_{p,i} &=  b\gamma_r \rho_{r,i} - \gamma_p \rho_{p,i}. \label{eq:rate-eqs}
\end{eqnarray}

The total local density we denote by $n_i$, and the cell-specific dephasing rate is given by:
\begin{eqnarray}
	\gamma_i  = \gamma_0 + \widetilde{C}_3 V_\mathrm{cell} \sum_{j\neq i} \frac{n_j\rho_{p,j}}{r_{ij}^{~3}} + \gamma_{ii} \nonumber
\end{eqnarray}
The first term, $\gamma_0$, is the dephasing due to other experimental factors, and the second and third term describe the dephasing due to interactions with other cells, and within one cell, respectively. The distance between cells $i$ and $j$ is denoted by $r_{ij}$, and $V_\mathrm{cell}$ is the physical volume of a cell. If the lattice spacing is $z_s$ (see figure~\ref{fig:honeycomb-calculation-lattice}), this equals $\sqrt{3}z_s^{~3}/2$.

The intracell interaction $\gamma_{ii}$ is obtained by integrating (\ref{eq:dipdip-interaction-integral}) from the typical distance between two atoms in the $\ket{r}$ state [$r_\mathrm{min}=\left(3/4\pi n\rho_r \right)^{1/3}$] to the typical radius associated with the cell size [$r_\mathrm{max}=\left( 3V_\mathrm{cell}/4\pi \right)^{1/3}$]. For $r_\mathrm{min}<r_\mathrm{max}$ this yields
\begin{eqnarray}
	\gamma_{ii}= \frac{4\pi}{3}\widetilde{C}_3 n_i \rho_{p,i} \ln\left(n \rho_{r,i} V_\mathrm{cell}\right). \nonumber
\end{eqnarray}
For $r_\mathrm{min}\geq r_\mathrm{max}$ we take $\gamma_{ii}=0$. Physically, this means there is only intracell dephasing when the blockade radius is smaller than the cell size. This is reasonable, since there is at most one Rydberg atom within a cell if the blockade radius is larger than that cell, which means that the dipolar interaction between atoms within the cell can be neglected. In practice, we find that $\gamma_{ii}$ only contributes very little to the calculated results, because the majority of cells is smaller than the blockade radius. The weak dependence on $\gamma_{ii}$ also confirms the dominance of the long-range (intercell) part of the dipolar interactions.

To keep the bookkeeping of the calculation manageable, we divide the cloud into slices along the major axis of the ellipsoid, and define a five-ring honeycomb lattice on the individual slices (see figure~\ref{fig:honeycomb-calculation-lattice}). Using the sixfold rotational symmetry, it is possible to identify groups of cells that have the same density and the same neighbors (up to a rotation). Combined with the reflection symmetry along the major axis as well, this means that we only need to calculate the populations for about one in twelve cells, provided we correctly account for the degeneracies in all the relevant terms.

We take a base dephasing rate $\gamma_0/2\pi = 1 \times 10^6$~s$^{-1}$, based on our previous experiments \cite{Cisternas17,deHond18}, and choose a lattice spacing $z_s = 1.5~\mathrm{\mu m}$ such that we have $\sim\!6500$ cells in total. This is approximately equal to the blockade radius ($\sim\!\mathrm{\mu m}$). It is a trade-off between the parts of the cloud that we cover, and the interactions that can be accounted for. Reducing the cell size means the description of the interaction should become more accurate, but at the same time we would need more cells to cover the cloud, which is computationally costly. In total, however, we see at most a few hundred atoms in the $\ket{p}$ state at any given time, meaning the average number per cell is much smaller than one.

\begin{figure}
	\centering
	\includegraphics{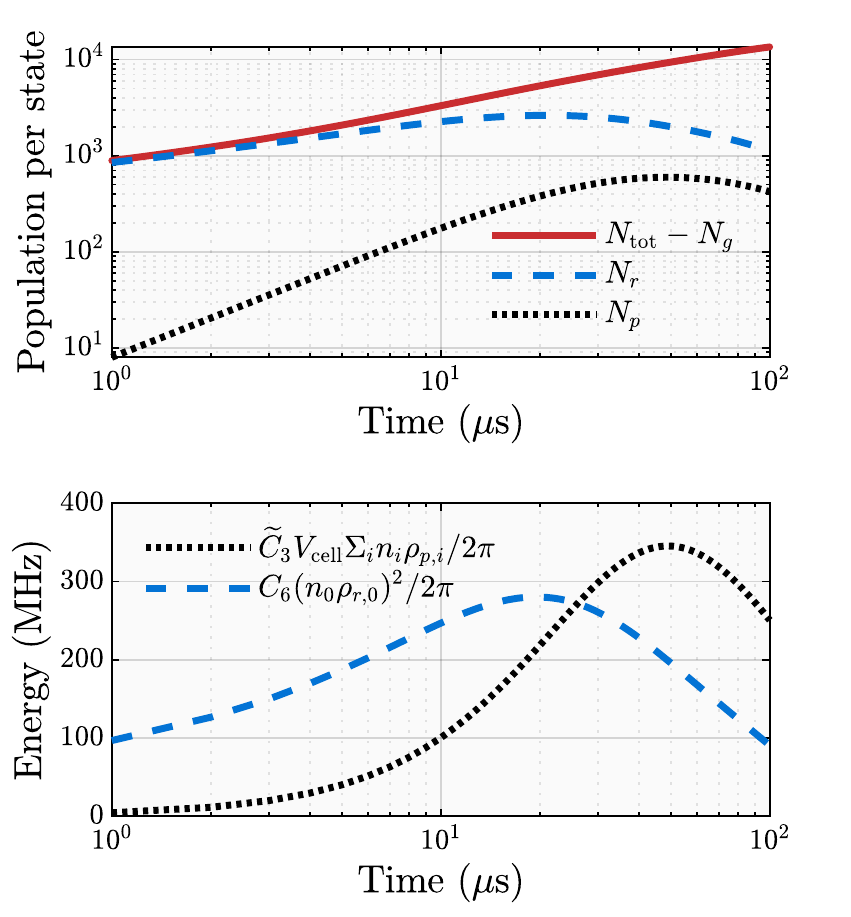}
	\caption{Resonant ($\Delta = 0$) time evolution in the cellular model of some relevant properties in the central cell at a peak density of $4~\mathrm{\mu m}^{-3}$ assuming a Rabi frequency of $2\pi \times 0.42~\mathrm{MHz}$. Top: population dynamics for a total atom number of $\sim\!20\times 10^3$. Bottom: typical energy scales for the dipole--dipole and van der Waals interactions. The energy scale associated with the van der Waals interaction is purely local, while the energy scale associated with the dipole--dipole interaction depends on all other cells, and hence contains a summation. Both energy scales exceed the linewidth ($\Gamma\approx 2\pi\times 50~\mathrm{MHz}$) observed in the spectrum for this Rabi frequency in figure~\ref{fig:cascade-plot} because they are computed for the central cell, which has a high density.}
	\label{fig:characteristicproperties}
\end{figure}
To obtain the dynamics of the total populations across the cloud, we take sums of the population fractions $\rho_{s,i}$ weighted by the local densities $n_i$. The coarseness of our lattice leads to some deviations between the experimental and the calculated atom number; to adjust for that, the simulated results are normalized to have the same baseline atom number (this is typically a $10\%$ correction).

We have simulated our measured spectra in this fashion, see figure~\ref{fig:cascade-plot}. They show a clear deviation from the non-interacting case, which is dominated by broadening due to depletion. We have extracted the linewidth and loss rate from these simulated features by fitting Gaussian lineshapes to them. These numbers have been included in figure~\ref{fig:linewidth-features} as well, where they show the same trend as the values we have found experimentally.

\begin{figure}
	\centering
	\includegraphics{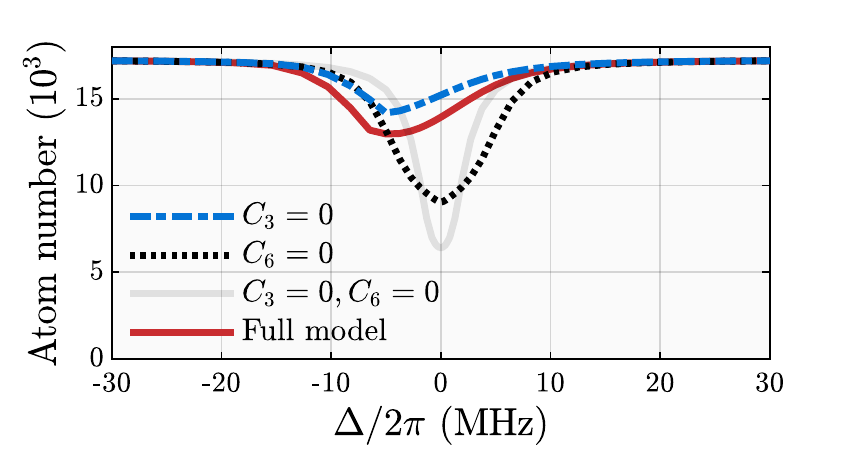}
	\caption{Spectra calculated from the cellular model, with parameters chosen to make the contributions of the different interactions stand out. For a pulse time of $10~\mathrm{\mu s}$ and a Rabi frequency of $2\pi\times0.42~\mathrm{MHz}$ it is clear that the van der Waals interaction by itself (the case $C_3 = 0$; blue, dash-dotted line) causes a strong suppression of losses when compared to the noninteracting case ($C_3 = 0, C_6 = 0$; light-gray line). On the other hand, when we look at the individual effect of the dipole--dipole interaction ($C_6 = 0$; dotted line) we mostly see a broadening. When combined in the fully-interacting case (red line) a broadened, suppressed spectrum emerges.}
	\label{fig:theoretical-spectrum}
\end{figure}

Our measured data show a slight asymmetric tail towards more negative detunings; this is typically ascribed to the excitation of Rydberg pairs at a distance smaller than the blockade radius~\cite{Reinhard08}, such that their energies are shifted. Our present model includes the interaction as a mean-field-like shift and thus does not take these pair states into account fully, and the asymmetric tail towards more negative detuning is not reflected in the results of our calculations. For certain parameters our model actually shows an asymmetric tail towards  \emph{less} negative detuning instead (also see figure~\ref{fig:theoretical-spectrum}). This happens because the mean-field term in (\ref{eq:rate-eqs}) itself is not constant, but depends on the Rydberg population. Thus the mean-field shift becomes detuning-dependent, which skews an otherwise Lorentzian lineshape in this manner.

In order to gain further insight, it is useful to consider the steady-state limit as well, as in Refs.~\cite{Goldschmidt16,Boulier17,Young17}.
The populations in the levels we are concerned with reach a steady state if we assume the system to be closed. The solution simplifies considerably if we furthermore assume the van der Waals interaction to be zero. This is not necessarily an accurate approximation for our Rydberg state of choice, but it will serve as a useful comparison because we can isolate the role of the dipole--dipole interaction. 

Under these assumptions, the resonant excitation rate is trivially $\Omega^2/\gamma$.
All populations are constant by definition, and they are fully determined by the excitation rate, lifetimes, and branching ratios.
In this regime, the contribution to the linewidth from the dipole--dipole interaction is typically tens of MHz, and can be assumed to be dominant over other experimental factors as captured by $\gamma_0$, which means the full dephasing rate can be approximated by \cite{Goldschmidt16,Boulier17}
\begin{eqnarray}
	\gamma \approx \gamma_{SS} = \sum_{i} \left| C_3^{~(i)} \right| n_0 \rho_{i}, \nonumber
\end{eqnarray}
where the summation is over the pollutant states such as in table~\ref{tab:bbr-culprits}. Here we have \cite{Goldschmidt16}
\begin{eqnarray}
	\rho_{i} = R \rho_{g} b_i  \tau_i, \nonumber
\end{eqnarray}
which we can use to express $\gamma_{SS}$ as
\begin{eqnarray}\label{eq:width-scaling}
	\gamma_{SS} = \Omega \sqrt{n_0 \rho_{g} \beta_3},
\end{eqnarray}
with
\begin{eqnarray}
	\beta_3 \equiv \widetilde{C}_3 \sum_{i}  b_i \tau_i. \nonumber
\end{eqnarray}
When summing over the four nearest states (see table~\ref{tab:bbr-culprits}) we find $\beta_3 = 4553~\mathrm{\mu m^3}$.
The results of this steady-state analysis are also indicated in figure~\ref{fig:linewidth-features}.

Using the relations given above we can derive the excitation rate to be $R = \Omega/\sqrt{n_0 \rho_{gg} \beta_3}$, but this significantly overestimates the loss rate, as can be seen in figure~\ref{fig:linewidth-features} (where we used a peak density of $n_0 = 4~\mathrm{\mu m}^{-3}$). The reason is that the van der Waals interaction manifests itself as a suppression of the Rydberg population, and hence losses, which element is missing in the above steady-state analysis. Thus, in the experiment the excitation rate slows down once a few excitations have been created.

The linewidth that we measure, on the other hand, is described rather well by (\ref{eq:width-scaling}). It may be a surprise that we can describe this without accounting for the blockade, but as (\ref{eq:rate-eqs}) shows, in our description the blockade only leads to an effective \emph{detuning}.

To obtain an intuition for the energy scales of the interactions, we can look at their time evolutions as given by the model, see figure~\ref{fig:characteristicproperties}.
In particular this shows that the characteristic energy scale associated with the dipole--dipole interaction is comparable to the van der Waals interaction after a few tens of $\mu$s, and dominates for longer timescales, which is where most of our data was obtained.

Motivated by this evolution, we can go to a slightly shorter timescale where the van der Waals interaction is still dominant, and look at the relative effects of the interactions. For a pulse of $10~\mathrm{\mu s}$, the calculated spectra clearly show that the van der Waals interaction suppresses the excitation, while the dipole--dipole interaction causes broadening (see figure~\ref{fig:theoretical-spectrum}).

From the point where the excited state populations saturate, the frequency associated with the dipolar exchange is so large, that a single pollutant will get a chance to explore the entire system within a Rydberg lifetime, as long as there are Rydberg states available to exchange with \cite{Gunter13}. This does not have consequences for the relative populations, so we are rather insensitive to it, but it is potentially interesting for experiments looking to study transport phenomena in strongly interacting systems, especially in the presence of decoherence \cite{Schonleber15}.

Recently, for instance, it has been argued that Rydberg-based, open quantum systems can be mapped onto simple models describing epidemic dynamics \cite{Perez-Espigares17,Grassberger83}. Such models describe the evolution of `individuals' between any of three states: susceptible, infectious, and recovered (SIR); in our terminology this would be $\ket{g}$, $\ket{r}$, and $\ket{l}$, respectively. By adding impurities it is possible to extend such a model to one that would include a generalized `incubation period' \cite{Li99}. This would yield a (quantum mechanical) analog of a Susceptible-Exposed-Infectious-Recovered (SEIR)-like model. In that case $\ket{r}$ and $\ket{p}$ would take the role of the exposed and infectious states, respectively.

In conclusion, we have studied the combined effects of the van der Waals interaction and resonant dipole--dipole interaction in experiments with Rydberg excitation in a dense and ultracold cloud of rubidium atoms, for timescales that are similar to the Rydberg state lifetime. The initial degree of Rydberg excitation is dominated by the van der Waals interaction. For later times, the dipole--dipole interaction dominates due to BBR-induced decay to nearby states. We have derived a cellular rate-equation model that includes both interactions, and takes the long-range character of the dipolar interaction into account. It is a first attempt to do so in a computationally tractable manner. These results indicate that the atomic cloud acts as a strongly interacting, driven-dissipative many-body system with an interplay between the (short-range) van der Waals interaction and the long-range dipolar interaction. They point to several interesting directions for further study.

\ack
We thank Shannon Whitlock for fruitful discussions. R.J.C.S.\ and N.J.v.D.\ acknowledge stimulating discussions within the QuSoft consortium. This work is part of the research programme of the Foundation for Fundamental Research on Matter (FOM), which is part of the Dutch Organisation for Scientific Research (NWO). We also acknowledge EU funding through the RySQ programme.

\section*{References}
\bibliography{bib}

\end{document}